\documentclass[%
reprint,
superscriptaddress,
 amsmath,amssymb,
 aps,
]{revtex4-2}

\usepackage{graphicx}% Include figure files
\usepackage{dcolumn}% Align table columns on decimal point
\usepackage{bm}% bold math
\usepackage{subfigure}
\usepackage[dvipsnames]{xcolor}
	
\usepackage{color, colortbl}
	
\definecolor{Gray}{gray}{0.9}
\definecolor{LightCyan}{rgb}{0.88,1,1}

\begin{document}

%\preprint{APS/123-QED}

\title{Correlated noise enhances
coherence and fidelity in coupled qubits}% Force line breaks with \\

\author{Eric~R.~Bittner}
\affiliation{Department of Chemistry, University of Houston, Houston, Texas 77204, United~States.}
\affiliation{Center for Nonlinear Studies, Los Alamos National Laboratory,
Los Alamos, NM 87545, United States}
\email{ebittner@central.uh.edu}

\author{Hao Li}
\affiliation{Department of Chemistry, University of Houston, Houston, Texas 77204, United~States.}

\author{S. A. Shah}
\affiliation{Center for Nonlinear Studies, Los Alamos National Laboratory,
Los Alamos, NM 87545, United States}
\affiliation{Theoretical Division, Los Alamos National Laboratory,
Los Alamos, NM 87545, United States}

\author{Carlos Silva}
\affiliation{School of Chemistry and Biochemistry, Georgia Institute of Technology, Atlanta, GA 30332, United~States}
\affiliation{School of Physics, Georgia Institute of Technology, Atlanta, GA 30332, United~States}
\affiliation{School of Materials Science and Engineering, Georgia Institute of Technology, Atlanta, GA 30332, United~States}

\author{Andrei Piryatinski}
\affiliation{Theoretical Division, Los Alamos National Laboratory,
Los Alamos, NM 87545, United States}

\date{\today}% 
\begin{abstract}
It is generally assumed that
environmental noise arising from thermal
fluctuations is detrimental to preserving
coherence and entanglement in a quantum
system.  In the simplest sense, dephasing and decoherence are tied to energy fluctuations driven
by coupling between the system and the normal
modes of the bath. 
Here, we explore the role of noise correlation in an open-loop model quantum communication system whereby
the ``sender'' and the ``receiver'' are subject to
local environments with various degrees of correlation or anticorrelation.  We introduce correlation
within the spectral density by solving a multidimensional stochastic differential equations
and introduce these into the Redfield equations of motion for the system density matrix. We find that correlation can enhance both the fidelity and purity
of a maximally entangled (Bell) state.  Moreover, we show that, by comparing the evolution of
different initial Bell states, one can
effectively probe the correlation between two local environments.  These observations may be useful
in the design of high-fidelity quantum gates and communication protocols. 
\end{abstract}

%\keywords{Suggested keywords}%Use showkeys class option if keyword
                              %display desired
\maketitle

\section{Introduction}
Noise and environmental fluctuations are generally considered detrimental
to the preservation of coherence and entanglement in an open quantum system.
Correlations between individual quantum
systems represent the basic resources in quantum information and quantum computing,
and one of the major technological tasks is to protect and control these correlations
and entanglements. Entanglement expresses the non-separability of the quantum state of a compound system. However, the coupling to environment leads to dissipation and loss of the quantum correlations, often on time scales much shorter than those needed for implementing quantum information tasks.
 Such fluctuations can arise from nuclear and electronic motions of the surrounding environment that induce a noisy driving field.
From the Anderson-Kubo (AK) model for spectral line-shapes
\cite{Kubo:JPSJ1954,Anderson:JPSJ1954,kubo1969stochastic,Mukamel1984},
the transition frequency obeys an Ornstein-Uhlenbeck (OU) process such that $\omega_t = \omega_o + \delta\omega_t$ and
\begin{align}
   d\delta\omega_t = -\gamma \delta\omega_t dt + \sigma dW_t
\end{align}
with $\langle \delta\omega(t)\rangle = 0$ and
\begin{align}
    \langle \delta\omega(t)\delta\omega(0)\rangle = \frac{\sigma^2}{2\gamma}e^{-\gamma |t|}
\end{align}
where $W_t$ is the Wiener process, $\Delta^2 = \frac{\sigma^2}{2\gamma}$ is the fluctuation amplitude, and ${1}/{\gamma} = \tau_c$ is the correlation time for the noise. According to the AK model, $\Delta\tau_c \ll 1$ corresponds to the case of fast modulation, which results in a purely Lorentzian spectral line shape and a pure dephasing time of $T_2 = (\Delta^2\tau_c)^{-1}$. Similarly, when $\Delta\tau_c \gg 1$, we are in the slow modulation regime and the spectral lineshape takes a purely Gaussian form reflecting the inhomogeneities of the environment.
We recently extended this approach 
to account for non-stationary/non-equilibrium
environments.
\cite{doi:10.1063/5.0026467,doi:10.1063/5.0026351}

Actively preventing decoherence from affecting quantum entanglement holds both theoretical and practical significance in quantum information processing technologies. 
Several recent studies suggest that
decoherence can be suppressed by
carefully engineering the system-bath 
coupling. 
\cite{Smirnov_2018,Golkar2018DynamicsOE,Hsiang2022EntanglementDO}
For example, Mouloudakis and Lambropoulos, extending previous work by Yang \textit{et al.}\cite{PhysRevA.79.012309}, studied the steady-state entanglement between two qubits that interacted asymmetrically with a common non-Markovian environment. The study found that, depending on the initial two-qubit state, the asymmetry in the couplings between each qubit and the non-Markovian environment could lead to enhanced entanglement in the steady state of the system.\cite{Mouloudakis2022CoalescenceON,Mouloudakis2021EntanglementII}

However, it is possible, especially in a condensed phase environment, that multiple modes of the environment can contribute to the frequency fluctuations, and it is possible that these contributions can be correlated, anti-correlated, or totally uncorrelated.
To set the stage for our subsequent analysis, let us consider what happens if we extend the AK model to account for multiple noise contributions. Consider a single stochastic process, $E_t$, described by a generalization of the Ito stochastic differential equation (SDE)\cite{Gardner},
\begin{align}
    dE_t = -\gamma E_t dt + B \cdot dW_t, \label{eq11}
\end{align}
where $B$ is a vector of variances $B = \{\sigma_1,\sigma_2\}$, and $dW = \{dW_1,dW_2\}$ are correlated Wiener processes with 
$dW_1(t)dW_2(t') = \delta(t-t') \xi dt$ and 
$dW_i(t)dW_i(t') = \delta(t-t') dt$,
where $-1 \leq \xi \leq 1$ is the correlation parameter between the two Wiener processes.  We can rewrite the SDE in Eq.~\ref{eq11} 
in terms of two uncorrelated processes by defining the variances
$B' = \{\sigma_1 + \sigma_2\xi, \sigma_2(1-\xi^2)^{1/2}\}$ 
such that Eq.~\ref{eq11} becomes
\begin{align}
    dE_t = -\gamma E_t dt + (\sigma_1+\sigma_2\xi)dW_1 + \sigma_2(1-\xi^2)^{1/2}dW_2
\end{align}
and $W_1$ and $W_2$ are now uncorrelated Wiener processes. If we work out the covariance of $E_t$ one finds that 
\begin{align}
    {Cov}[E_t,E_s] = \frac{e^{-\gamma (s+t)}}{2\gamma}(e^{2\gamma {\rm min}(s,t)}-1) \sigma_{\mathrm{eff}}^2
\end{align}
where we can define an effective covariance parameter
\begin{align}
    \sigma_{\mathrm{eff}}^2 = \sigma_1^2 + \sigma_1\sigma_2 \xi + \sigma_2^2.
\end{align}
We see from this that anticorrelation $(\xi < 0)$ leads to a net decrease in the covariance function for a given stochastic process. This implies that a system coupled to an anticorrelated environment will have a longer dephasing time as compared to a system coupled to uncorrelated or completely correlated baths. 

\section{Theory}
Our theory is initialized by assuming
that the total system can be separated into system and reservoir variables such that
\begin{align}
H = H_{o} + \sum_k \hat A_k E_k(t) = H_o + H_r(t)
\end{align}
where $H_{o}$ describes the system independent of
reservoir
with eigenstates $H_o|\alpha\rangle = \hbar\omega_\alpha|\alpha\rangle$, 
$\hat A_k$ are a set of
quantum operators acting on the system subspace, and $E_k(t)$ are stochastic variables representing the
dynamics of the environment.
Formally, we write these in 
terms of an Ito stochastic differential equation of the 
form 
\begin{align}
    d{\mathbf{E}} &= \mathbf{A}(\tau, \mathbf{E}(\tau) )dt
    + \mathbf{B}(\tau, \mathbf{E}_\tau )\cdot d\mathbf{W}_t
\end{align}
where $\mathbf{W}$ is a vector of Wiener processes and 
$\mathbf{A}(t,\mathbf{E})$ and $\mathbf{B}(t,\mathbf{E})$
define the 
the drift and the diffusion.
This general form allows for both nonlinear and geometric
processes to be incorporated
into our model on an even footing.
The process $\mathbf{E}(t)$ is in general multidimensional and driven by a multidimensional Wiener process with correlation matrix
$\Sigma$.  
The process itself can be 
written in integral form as
\begin{align}
    \mathbf{E}(t)-
    \mathbf{E}({t_o})
    &= 
    \int_{t_o}^t d\tau
    \mathbf{A}(\tau, \mathbf{E}(\tau) )\nonumber \\
    &+ \int_{t_o}^t d\tau
    \mathbf{B}(\tau, \mathbf{E}(\tau))\cdot d\mathbf{W}(t)
\end{align}
with $dW_i(t)dW_j(t') = \delta(t-t')\Sigma_{ij}dt$
as the generalized statement 
of  Ito's lemma. 
If we take the noise terms to be correlated Ornstein-Uhlenbeck 
processes with
\begin{align}
    d\mathbf{E}_t = -\mathbf{A}\cdot \mathbf{E}_t dt + \mathbf{B}\cdot d\mathbf{W}_t
\end{align} 
with $\mathbf{\Sigma}dt\delta(t-t') = d\mathbf{W}\otimes d\mathbf{W}$
as the correlation matrix, 
the general spectral density 
matrix takes the form
\begin{align}
    J(\omega) 
    = \frac{1}{2\pi}(\mathbf{A}+i\omega)^{-1}\cdot\mathbf{B}\cdot\mathbf{\Sigma}\cdot \mathbf{B}^T\cdot (\mathbf{A}-i\omega)^{-1},
\end{align}
as derived in Appendix~\ref{append:3} (c.f. Eq.~\ref{eq:90}).
These terms enter into the 
quantum dynamics of the
reduced density matrix for the 
system variables via the 
Bloch Redfield equations
\begin{align}
    d_t\rho_{\alpha\alpha'} = -i (\omega_\alpha-\omega_\alpha')\rho_{\alpha\alpha'}
    -\sum_{\beta\beta'}{\cal R}_{\alpha\alpha',\beta\beta'}(\rho_{\beta\beta'}-\rho_{\beta\beta'}^{eq})
\end{align}
where  $\rho^{eq}$ is the equilibrium reduced density matrix and ${\cal R} $ is the Bloch-Redfield tensor
\begin{align}
\begin{split}
     {\cal R}_{\alpha\alpha',\beta\beta'} 
    =
    \sum_{nm}\left\{
    \delta_{\alpha'\beta'}
    \sum_{\gamma}J_{nm}(\omega_\beta-\omega_\gamma)(A_n)_{\gamma\beta} (A_m)_{\alpha\gamma} \right.\\  
    \left.- (J_{nm}(\omega_\alpha'-\omega_\beta') + J_{nm}(\omega_\beta-\omega_\alpha))(A_n)_{\beta'\alpha'} (A_m)_{\alpha\beta} \right. \\
    +\left.\delta_{\alpha\beta}
    \sum_\gamma J_{nm}(\omega_\gamma-\omega_\beta')(A_n)_{\beta'\gamma} (A_m)_{\gamma \alpha'}
    \right\}
\end{split}
%\label{eq:5}
\end{align}
where 
$(A_n)_{\alpha\beta} = \langle\alpha | \hat A_n | \beta\rangle$ are the matrix elements of the $\hat A_n$ operator in the 
eigenbasis of $H_o$ and 
$J_{nm}(\omega)$ are elements of the generalized spectral matrix 
characterizing the coupling between the 
system and its environment. \cite{Konrat:1993aa,redfield1957theory,Solomon1955559,ArgyresKelley1964}

\begin{figure}[htbp]
    \centering
    \includegraphics[width=\columnwidth]{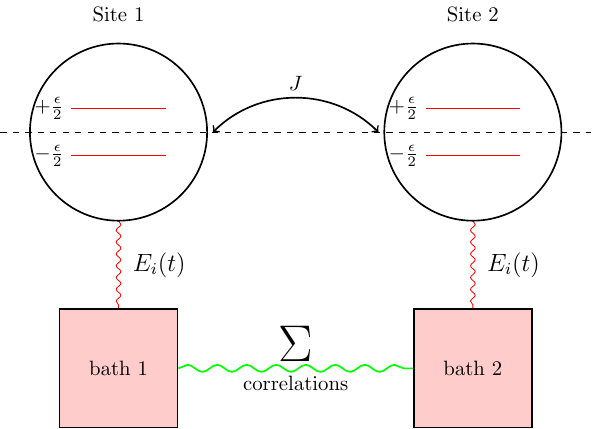}
    \caption{Sketch of 2-site model with correlated 
    noise interactions. }
    \label{fig:enter-label}
\end{figure}

Under the secular approximation 
in which the 
time evolution of the system is slow 
compared to the characteristic 
correlation time of the environment $|\omega_{\alpha\beta}-\omega_{\gamma\delta }| \ll 1/\tau_c$,
the population terms on the diagonal
can be decoupled from the off-diagaonal coherence terms
via
\begin{align}
    {\cal R}_{ij;kl}^{sec}
    = \delta_{ij} \delta_{kl}+\delta_{ik}
    \delta_{jl} (1-\delta_{ij} \delta_{kl}).
\end{align}
The time
evolution of the reduced density matrix
is strictly unitary under the secular approximation,
which guarantees that the ${\rm tr}(\rho) =1$
 and all diagonal elements are positive.  

Table ~\ref{tab:1} gives a list of
Redfield tensor elements for a single $SU(2)$ qubit
driven by correlated noise in both longitudinal ($\sigma_z$) and spin-lattice $(\sigma_x)$ terms. 
Within the secular approximation, the longitudinal
terms contribute to the pure dephasing $(T_2)$ time, while the spin-lattice term contributes to the
relaxation.  Even when
the diffusion matrix $\mathbf{B}$ is diagonal, cross-correlation enters the Redfield tensor
via non-vanishing terms involving the cross-spectral densities; however, 
these terms only contribute to the non-secular terms
of the tensor.

\subsection{Coherence transfer between two qubits}

We can easily generalise this model to accompany
any number of states to explore how correlated noise affects the
relaxation dynamics of the system. 
Here we consider a system of two spatially separated qubits, each
driven by locally correlated fields, coupled together by 
a static dipole-dipole interaction and coupled to 
environment 
\begin{align}
    H &= \sum_{i=1,2}\frac{\epsilon_i}{2}\hat\sigma_{i}(3) + J(\hat\sigma_1^+ \hat\sigma_2^- + \hat\sigma_2^+ \hat\sigma_1^-) + \sum_{j} \hat A_j E_j(t)
    \end{align}
where $E_j(t)$ are stochastic processes as above. 
Any system operators in the state space $SU(2)\otimes SU(2)$
can be constructed by taking the tensor products of $SU(2)$ Pauli matrices. 
Physically, this model could be achieved in systems in
which the energy of the local sites
are strongly modulated by the
local phonon modes, as in the case of
Jahn-Teller distortions of high-spin octahedral $d^4$ coordination
compounds
where axial or equatorial distortions split the
otherwise degenerate $d_{z^2}$ and $d_{x^2-y^2}$ orbitals.
Consequently, for a pair of octahedral sites, one can have
symmetric and antisymmetric combinations of normal modes that drive the Jahn-Teller distortions of
each site, giving rise to various degrees of correlation
of the thermal noise experienced at each metal site. 
Furthermore, it may be possible through chemical or external
stimulation to selectively enhance these modes.

% For
% example, the dipole operator is given by
% \begin{align}
%     \hat \mu &= \mu(\hat\sigma(0)\otimes \hat\sigma(1) + \hat\sigma(1)\otimes \hat\sigma(0)).
% \end{align}
% Similarly, we construct the operators coupling the system to the environment as 
% \begin{align}
% \begin{aligned}
% \hat A_{1}&= \hat \sigma(3)\otimes \hat\sigma(0) \\
% \hat A_{2}&= \hat \sigma(0)\otimes \hat \sigma(3) \\
% \hat A_{3}&= \hat \sigma(1)\otimes \hat \sigma(0) \\
% \hat A_{4}&= \hat \sigma(0)\otimes \hat \sigma(1),
% \end{aligned}
% \end{align}
% which for the uncorrelated case, would correspond to each spin being coupled 
% to the environment via longitudinal ($\hat A_1$ and $\hat A_2$) or transverse 
% ($\hat A_1$ and $\hat A_2$) components.   The tunneling operator can
% also be constructed as 
% \begin{align}
% (\hat\sigma_1^+ \hat\sigma_2^- +\hat \sigma_2^+ \hat\sigma_1^-) = \hat\sigma^+\otimes \hat\sigma^- + \hat\sigma^+\otimes \hat\sigma^-
% \end{align}
% where $\hat\sigma^\pm = (\hat\sigma(1)\pm i \hat\sigma(2))/2$.
If the energy states of the 
uncoupled qubits are identical, $\epsilon_1 = \epsilon_2$, then the two tunneling states are
symmetric and antisymmetric combinations of singly excited configurations
$|10\rangle$ and $|01\rangle$, that is,
 \begin{align}
     |\psi_\pm\rangle = \frac{1}{\sqrt{2}}
     (|01\rangle \pm |10\rangle )
 \end{align}
 and dipole transitions from the lowest energy $|00\rangle$ are only to the 
 symmetric linear combination.  If $J>0$ the symmetric state lies higher in energy
than the antisymmetric state and vice versa when $J<0$. For $\epsilon_1\ne \epsilon_2$ 
both states are optically coupled to the ground state, producing a pair of optical transitions, one of which being more intense than the other (superradiant vs. subradiant).
\begin{widetext}

We examine the effect of cross-correlation 
by computing the linear absorption spectrum of the system
for a suitable choice of parameters. 
From time-dependent 
perturbation theory, the linear absorption spectrum is given by
\begin{align}
    S(\omega) \propto \left|\frac{1}{i\hbar} \int_{-\infty}^\infty e^{i\omega t} \langle\mu(t)[\mu(0),\rho(-\infty)]]\rangle \right|^2
\end{align}
where $\hat \mu(t)$ is the transition dipole operator in the Heisenberg/Schr\"odinger representation at time $t$ and $\rho(-\infty)$ 
is the system density matrix at $t\to -\infty $.

\begin{figure*}[htbp]
\begin{center}
\subfigure[]{\includegraphics[width=0.4\columnwidth]{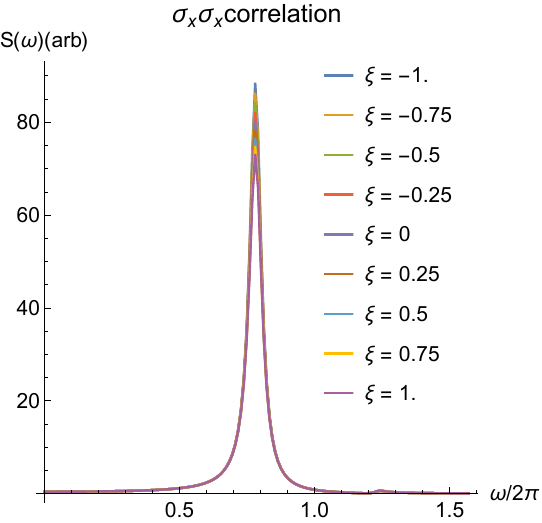}}
\subfigure[]{\includegraphics[width=0.4\columnwidth]{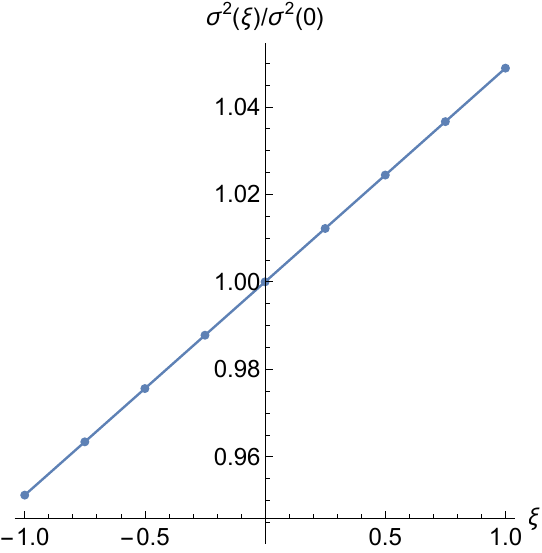}}
\subfigure[]{\includegraphics[width=0.4\columnwidth]{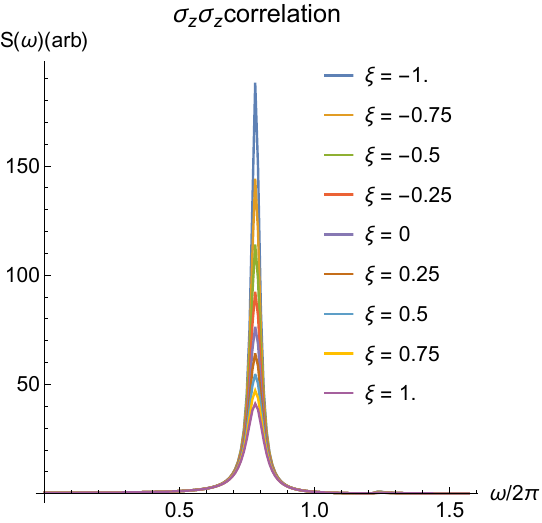}}
\subfigure[]{\includegraphics[width=0.4\columnwidth]{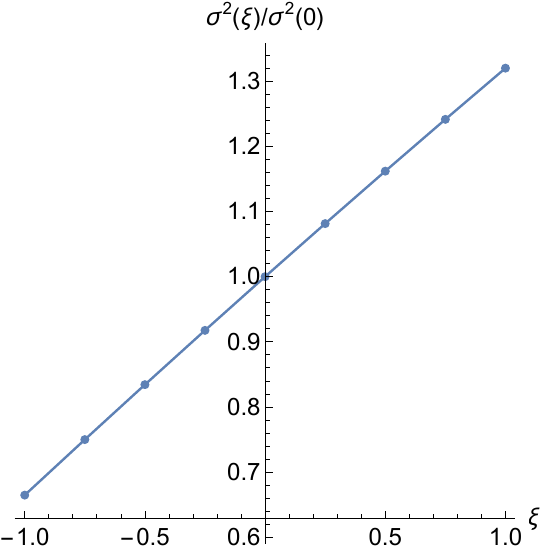}}
\caption{Linear response absorption spectra
and associated line-width for 
a pair of qubits subject to  transverse (a,b)
and longitudinal (c,d) noise terms with various
degrees of correlation. 
}
\label{figure:2}
\end{center}
\end{figure*}
Fig.~\ref{figure:2}(a-d) shows the linear absorption spectra
and the corresponding relative line widths for a
pair of qubits with interaction $J/\epsilon = -0.2$ and with
correlation between \textit{either} the two transverse or the
two longitudinal noise terms.  
Since only two noise terms
are correlated, the spectral density matrix is given by
\begin{align}
    J(z) =
\frac{1}{2\pi}
    \left(
\begin{array}{cc}
 \frac{2 \xi  \sigma _{12} \sigma _1+\sigma _1^2+\sigma _{12}^2}{\gamma _1^2+z^2} & \frac{\xi  \sigma _1 \sigma _2+\xi 
   \sigma _{12} \sigma _{21}+\sigma _{12} \sigma _2+\sigma _1 \sigma _{21}}{\left(z-i \gamma _1\right) \left(z+i \gamma
   _2\right)} \\
 \frac{\xi  \sigma _1 \sigma _2+\xi  \sigma _{12} \sigma _{21}+\sigma _{12} \sigma _2+\sigma _1 \sigma _{21}}{\left(z+i
   \gamma _1\right) \left(z-i \gamma _2\right)} & \frac{2 \xi  \sigma _{21} \sigma _2+\sigma _2^2+\sigma _{21}^2}{\gamma
   _2^2+z^2} \\
\end{array}
\right)
\end{align}
to denote whether the  term is local to site 1 or 2 or 
involves explicit correlation between the two. Again, $\xi$ denotes
whether or not the terms are correlated or anticorrelated. 
In the transverse-transverse case, $J(z)$ is evaluated at the transition frequency, since this
coupling involves the inelastic
coupling to the environment; where as in the longitudinal-longitudinal
case, $J(z)$ is evaluated at $z=0$ since this corresponds to
a purely elastic coupling between the system and the
environment.  

\end{widetext}

Here, we see that correlations between transverse components
have little effect on the spectral line shape.  
We can
understand this since the spectral density terms are all evaluated
at the transition frequency and are always smaller than their
longitudinal counterparts.  

On the other hand, the correlations
between longitudinal components have a much more dramatic effect
on both the transition intensity and line width, with
anti-correlated noise giving much sharper and more intense
transitions. We can understand this in the following way. 
According to the Kubo-Anderson model, the spectral lineshape is
determined by fluctuations in the transition frequency.  In the anticorrelated case, the local fluctuations are perfectly
synchronised but in opposite ways. That is, as
the local site energy of one increases, the other site energy
always decreases. Therefore, the two local fluctuations cancel each other
out.  In the fully correlated case, the fluctuations are also
perfectly synchronised, but both site energies increase or decrease, which results in a broader spectral transition.

\begin{figure*}[hbtp]
\begin{center}

\subfigure[]{\includegraphics[width=\columnwidth]{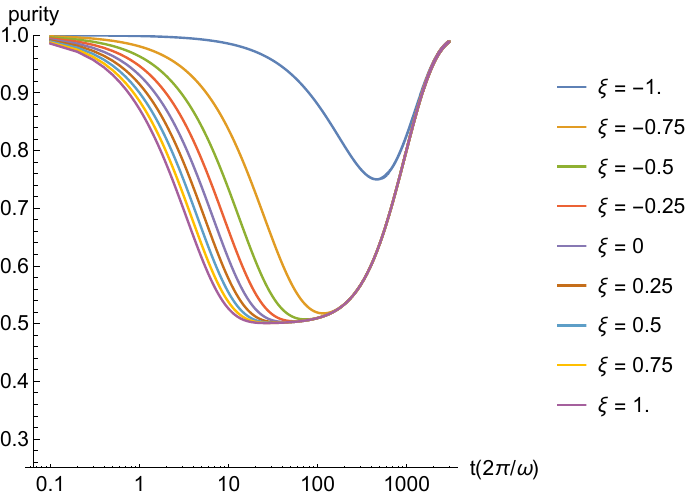}}
\subfigure[]{\includegraphics[width=\columnwidth]{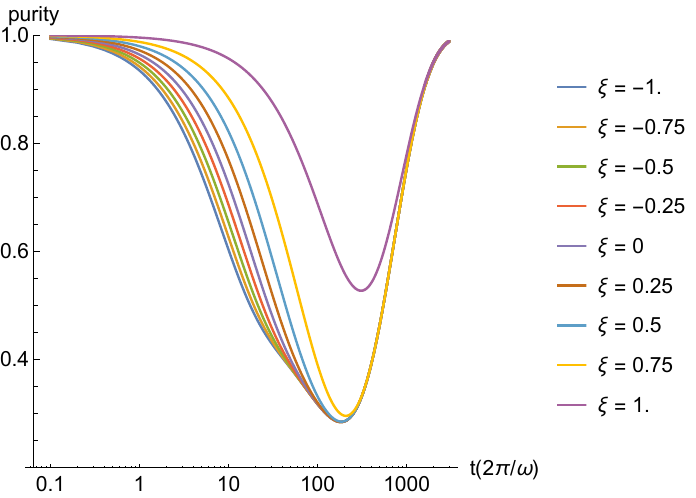}}
\subfigure[]{\includegraphics[width=\columnwidth]{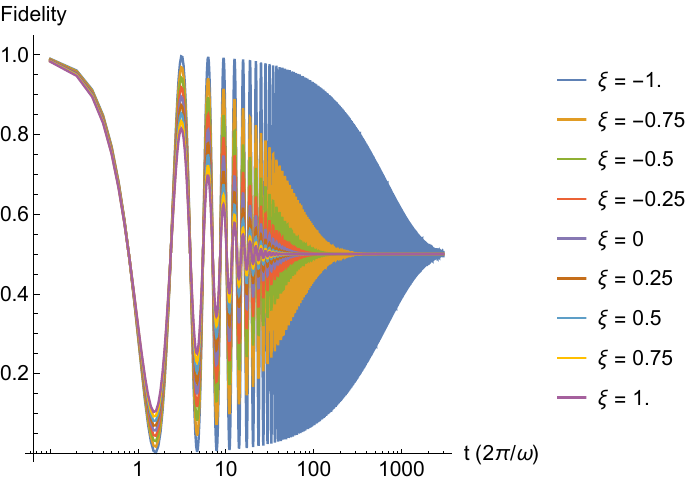}}
\subfigure[]{\includegraphics[width=\columnwidth]{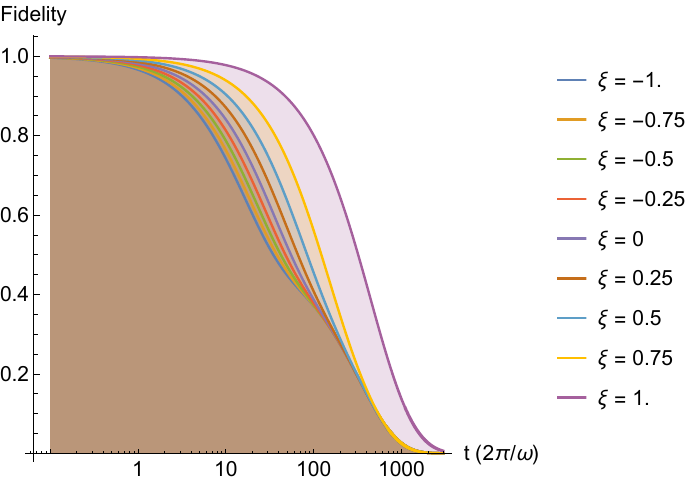}}
\caption{Purity(a,b) and fidelity (c,d) of composite $SU(2)\otimes SU(2)$ qubit pair
vs time for systems prepared in a  maximally entangled (Bell) state
corresponding to (a,c)$|\Phi^+\rangle$ or  (b,d) $\Psi^+$.}
\label{fig:3}
\end{center}
\end{figure*}

We next consider the effect of initial-state preparation on
the quantum dynamics of the entangled qubits. 
For this, we introduce the following four Bell-states
\begin{align}
    |\Phi^{\pm}\rangle &= \frac{1}{2}(|00\rangle \pm |11\rangle)\\
    |\Psi^{\pm}\rangle &= \frac{1}{2}(|01\rangle \pm |10\rangle)
\end{align}
which correspond to the four maximally  
 entangled quantum states of two qubits. 
From the previous discussion, longitudinal correlations
appear to have the most profound effect on the dynamics,
so we shall consider only that sort of coupling in this
example. 

The purity, $\gamma = {\rm tr}(\rho^2)$, provides a 
useful measure of the degree to which a quantum state
is mixed. Mathematically, $\gamma = 1$ for a pure state 
since $\rho = \rho^2$ 
and takes a lower bound of $\gamma = 1/4$ corresponding to
the case where all 4 states of the system are equally probable. 
Initially, the system is in a pure state with $\gamma =1$
and evolves toward a mixed state as it evolves.  
At long time and low temperature, the system will relax completely to the ground state $|00\rangle$ with $\gamma= 1$.

Fig.~\ref{fig:3}(a,b) shows the 
purity of a composite $SU(2)\otimes SU(2)$ qubit pair
vs time for systems prepared in a maximally entangled (Bell) state
and subject to longitudinal noise with various degrees 
of correlation or anti-correlation. 
In Fig.~\ref{fig:3}(a), we take the
initial state as a coherence between
the doubly excited state $|11\rangle$ and  the ground state $|00\rangle$, corresponding to the $\Phi_{+}$ Bell state.
Here, anticorrelation leads to a profound increase in
the system's ability to retain its purity for nearly two orders of
magnitude in time longer than the fully correlated case. 
In contrast, if the initial state is prepared in one of the
$\Psi_\pm$ Bell states, corresponding to a linear combination within
the singly excited manifold of states,  correlation {\em enhances} 
the systems ability to retain purity. The only difference between
the two results is in the preparation of the initial state. 
This provides a potentially useful experimental means for
determining the correlation or  anti-correlation between 
local environments. 

However, the fidelity 
\begin{align}
    F(\rho,\sigma)
     = 
     \left(
     {\rm tr}\left(\sqrt{
     \sqrt{\rho}
     \sigma \sqrt{\rho}
     }\right)
     \right)^2
\end{align}
is also an important 
consideration for whether or not a given Bell state is suitable for an shared key. 
Fidelity provides a measure of the ``closeness" of two quantum states. It expresses the probability that one state will pass a test to identify itself as the other.  By symmetry, $F(\rho,\sigma) = F(\sigma,\rho)$.   In Fig.~\ref{fig:3}(c,d), we compute the time-evolved fidelity $F(\rho_0,\rho_t)$ starting
from the $\Phi^+$(c) or
$\Psi^+$(d) Bell states, versus
various degrees of correlation
between the baths.  
For $\Phi^+$, the system loses fidelity rapidly and undergoes Rabi
oscillation within the double excitation manifold spanned by $\Phi^+$
and $\Phi^-$. The fidelity
eventually relaxes to $F =1/2$ for a long time, corresponding to complete relaxation into the ground state $|00\rangle$. 
As with purity, the envelope
of fidelity is enhanced
by anti-correlated noise. 

In contrast, the {\em correlated} noise helps to maintain both the
purity and fidelity of the
$\Psi^+$ Bell state.  This
state is an eigenstate of the
bare system Hamiltonian which can be prepared by direct photoexcitation from the ground state. It is curious that the above results suggest that correlated noise suppresses the optical response. However, 
the optical response is actually a measure of the coherence between the ground state and $\Psi^+$ and not
a measure of the
purity or fidelity of a given
state.

\section{Discussion}
In this paper we explored the role of noise correlation on a model open
quantum system consisting of one and two coupled $SU(2)$ qubits and showed how the dynamics and spectroscopy of the system
can be profoundly affected by environmental correlations. 
This has deep implications for searching materials suitable for quantum communications and computation applications in which long coherence times
and retention of are required.  
In the case of super-dense coding,
a sender (A) and a receiver (B)
 can communicate a number of classical bits of information by only transmitting a smaller number of qubits, provided that A and B
are presharing an entangled resource. \cite{PhysRevLett.69.2881} 
 Since this resource is subject to environmental noise, the ability of A and B to perform super-dense coding
hinges on their ability to maintain the fidelity of the state of the shared resource.
Similarly, quantum teleportation
requires the sender and receiver to
share a maximally entangled state.
\cite{PhysRevLett.70.1895} 
Our results suggest that by knowing whether the state is subject to correlated or anticorrelated noise, A and B
can be ensured that their shared resource state can maintain its purity
long enough for the information to be communicated. 
We also suggest that the
local noise correlation can be tuned by manipulating the local environment
around the two qubits.

\begin{acknowledgments}
The work at the University of Houston was funded in
part by the National Science Foundation (CHE-2102506) and the Robert A. Welch Foundation (E-1337). 
The work at Los Alamos National Laboratory was funded by the Laboratory Directed Research and Development (LDRD) programme, 20220047DR.
%ERB-Ulam LUR number?
The work at Georgia Tech was funded by the National Science Foundation (DMR-1904293). %
\end{acknowledgments}
%\tableofcontents
%\nocite{*}

\section*{Author Contributions}
\textbf{Eric Bittner:} Supervision, 
Funding acquisition,
Conceptualisation, Methodology, Formal analysis, Validation, Writing;
\textbf{Hao Li:} Formal analysis, Methodology, Validation;
\textbf{Syad A Shah: }Conceptualisation;
\textbf{Carlos Silva:} Conceptualisation
Funding acquisition;
\textbf{Andrei Piryatinski: }Conceptualisation,  Validation, Funding acquisition.
All authors contributed to the final draft 
and editing of this manuscript. 

\section*{Data Availability}

Data supporting the findings of this study are available from the corresponding author on a reasonable request.

%\bibliography{Redfield-notes}% Produces the bibliography via BibTeX.
%apsrev4-2.bst 2019-01-14 (MD) hand-edited version of apsrev4-1.bst
%Control: key (0)
%Control: author (8) initials jnrlst
%Control: editor formatted (1) identically to author
%Control: production of article title (0) allowed
%Control: page (0) single
%Control: year (1) truncated
%Control: production of eprint (0) enabled
%

\appendix
\begin{widetext}
% !TEX root=Noise-enhanced-fidelity.tex

\section{Correlations amongst random variables}
\label{append:3}

The Ornstein-Uhlenbeck process is a very useful method to account for many Markovian stochastic processes. Its multivariate representation is even more practical for physical processes. Here we discuss the multivariate Ornstein-Uhlenbeck process, including correlated Wiener processes, for the purpose of tackling realistic physical problems such as chromorphores coupled to their respective phonon environments but interacting with a common bath.

We write the multivariate Ornstein-Uhlenbeck process as a vector ${\bm E}(t)$ composed of individual processes $X_i(t)$. The stochastic differential equation reads
\begin{equation}
    {\rm d} {\bm E}(t) = A \left[ {\bm \mu} - {\bm E}(t)\right] {\rm d}t + B {\rm d} {\bm W}(t),
    \label{eqn:MOU-SDE}
\end{equation}
in which $A$ and $B$ are coefficient matrices, ${\bm \mu}$ is the vector of the Wiener process drift $\mu_i$ corresponding to $W_i$. ${\bm W}(t)$ is the vector of Wiener processes $W_i(t)$ which are correlated through the correlation matrix
\begin{equation}
    \xi(t,t') \equiv \delta_{tt'} {\rm d}{\bm W}(t){\rm d}{\bm W}(t')^{\rm T} /{\rm d}t,
    \label{eqn:correlation_matrix}
\end{equation}
where the angular brackets represent the ensemble average. The matrix elements $\xi_{ij}={\rm d}W_i(t) {\rm d}W_j(t)/{\rm d}t$ are defined through the It{\^o} isometry in higher dimensions. Obviously $\xi_{ii}=1$ according to the quadratic variation $({\rm d} W_t)^2={\rm d}t$. $\xi_{ij}$ varies from -1 to 1, respectively, corresponding to the fully anticorrelated and fully correlated cases. $\xi_{ij}=0$ means that the two Wiener processes are completely uncorrelated.

According to the It{\^ o}'s lemma, one finds the solution
\begin{equation}
    {\bm E}(t) = e^{-At} {\bm E}(0) + (\mathbb{1} - e^{-At}) {\bm \mu} + \int_0^t e^{-A(t-t')} B {\rm d}{\bm W}(t'),
    \label{eqn:MOU-solution}
\end{equation}
where ${\bm E}(0)$ is the initial condition of the process ${\bm E}(t)$, the mean value
\begin{equation}
    \langle {\bm E}(t) \rangle = e^{-At} \langle {\bm E}(0)\rangle + (\mathbb{1} - e^{-At}) {\bm \mu},
    \label{eqn:MOU-mean}
\end{equation}
and the correlation function
\begin{align}
    \nonumber \left< {\bm E}(t),{\bm E}^{\rm T}(s) \right> &\equiv \left< \left[{\bm E}(t) - \langle{\bm E}(t)\rangle\right]\left[{\bm E}(s) - \langle{\bm E}(s)\rangle\right]^{\rm T} \right> \\
    &= e^{-At}\left<{\bm E}(0),{\bm E}^{\rm T}(0)\right> e^{-A^{\rm T}s} + \int_0^{\min(s,t)} e^{-A(t-t')}B\xi B^{\rm T}e^{-A^{\rm T}(s-t')} {\rm d}t'
    \label{eqn:MOU-correlation}
\end{align}
following the It{\^ o} isometry in higher dimensions.

If $AA^{\rm T}=A^{\rm T}A$, one can find a unitary matrix $S$ to diagonalize the coefficient matrix $SAS^{\dagger}=SA^{\rm T}S^{\dagger}={\rm diag}(\gamma_1,\gamma_2,\dots,\gamma_n)$. For deterministic initial condition $\left<{\bm E}(0),{\bm E}^{\rm T}(0)\right>=0$, so does the correlation function $\left< {\bm E}(t),{\bm E}^{\rm T}(s) \right> = S^{\dagger}G(t,s)S$, in which
\begin{align}
    \begin{split}
        \left[G(t,s)\right]_{ij} &= \frac{\left(B\xi B^{\rm T}\right)_{ij}}{\gamma_i+\gamma_j}\left[e^{-\gamma_i|t-s|} - e^{-\gamma_it-\gamma_js}\right] \qquad (t\geq s), \\
        \left[G(t,s)\right]_{ij} &= \frac{\left(B\xi B^{\rm T}\right)_{ij}}{\gamma_i+\gamma_j}\left[e^{-\gamma_j|t-s|} - e^{-\gamma_it-\gamma_js}\right] \qquad (t\leq s).
    \end{split}
    \label{eqn:correlation-element}
\end{align}

If the real parts of all $A$'s eigenvalues are positive, one finds the stationary solution
\begin{equation}
    {\bm E}_{\rm s}(t) = {\bm \mu} + \int_{-\infty}^t e^{-A(t-t')}B{\rm d}{\bm W}(t'),
\end{equation}
with the stationary correlation matrix
\begin{equation}
    \left< {\bm E}_{\rm s}(t),{\bm E}_{\rm s}^{\rm T}(s) \right> = \int_{-\infty}^{\min(s,t)} e^{-A(t-t')}B\xi B^{\rm T} e^{-A^{\rm T}(s-t')}{\rm d}{\bm W}(t').
    \label{eqn:stationary-correlation}
\end{equation}

We define the stationary covariance matrix
\begin{equation}
    \sigma=\left< {\bm E}_{\rm s}(t),{\bm E}_{\rm s}^{\rm T}(t) \right>,
\end{equation}
then find a useful algebraic equation for stationary covariance matrix
\begin{equation}
    A\sigma +\sigma A^{\rm T} = B\xi B^{\rm T}.
    \label{eqn:stationary-covariance-algebra}
\end{equation}
For $s<t$ the stationary correlation function Eq.(\ref{eqn:stationary-correlation}) can be written as
\begin{align}
    \left< {\bm E}_{\rm s}(t),{\bm E}_{\rm s}^{\rm T}(s) \right> &= e^{-A(t-s)} \int_{-\infty}^s e^{-A(s-t')}B\xi B^{\rm T} e^{-A^{\rm T}(s-t')}{\rm d}t' \nonumber \\
    &=e^{-A(t-s)}\sigma \qquad s<t,
\end{align}
and 
\begin{equation}
    = \sigma e^{-A^{\rm T}(s-t)} \qquad s>t.
\end{equation}
The correlation function only depends on the time difference $|t-s|$ as expected for the stationary solution.
%Note that for stationary solution $\sigma(t)=\sigma(s)=\sigma$.
We define the stationary correlation matrix $G_{\rm s}(\tau)=\left< {\bm E}_{\rm s}(t),{\bm E}_{\rm s}^{\rm T}(t-\tau) \right>$, obviously $G_{\rm s}(0)=\sigma$. Then the above relation can be written in the form of the regression theorem 
\begin{align}
    \frac{\rm d}{\rm d\tau} \left[G_{\rm s}(\tau)\right] = \frac{\rm d}{\rm d\tau} \left<{\bm E}_{\rm s}(\tau),{\bm E}_{\rm s}^{\rm T}(0) \right> = -A G_{\rm s}(\tau).
\end{align}
Noting that $G_{\rm s}(0)=\sigma$, one can compute the stationary correlation matrix.

Since $\sigma^{\rm T}=\sigma$, we have
\begin{equation}
    G_{\rm s}(\tau) = \left[G_{\rm s}(-\tau)\right]^{\rm T}.
\end{equation}
Therefore, one can find the spectrum matrix as the Fourier transform of the autocorrelation matrix $G_{\rm s}(\tau)$
\begin{align}
    J(\omega) &= \frac{1}{2\pi} \int_{-\infty}^{\infty} e^{-i\omega \tau} G_{\rm s}(\tau) {\rm d}\tau \nonumber \\
    &= \frac{1}{2\pi} \left(A+i\omega\right)^{-1} B\xi B^{\rm T} \left(A-i\omega\right)^{-1}.
\label{eq:90}
\end{align}

As an example, we consider the case of the case of two correlated modes, in which 
we define the 2D Ornstein-Uhlenbeck process by the SDEs
\begin{align}
    {\rm d}E_1(t) &= -\gamma_1 E_1(t) {\rm d}t + \sigma_{11} {\rm d}B_1(t) + \sigma_{12} {\rm d}B_2(t), \nonumber \\
    {\rm d}E_2(t) &= -\gamma_2 E_2(t) {\rm d}t + \sigma_{21} {\rm d}B_1(t) + \sigma_{22} {\rm d}B_2(t). \nonumber
\end{align}
The two Wiener processes $B_1(t)$ and $B_2(t)$ are coupled through the correlation parameter $\xi={\rm d}B_1(t){\rm d}B_2(t)/{\rm d}t$. The range of $\xi$ is between $-1$ to $1$ corresponding to the cases of complete anti-correlation and correlation, respectively. $\xi=0$ means that the two Wiener processes are completely decoupled.
The solutions of the OU processes are
\begin{align}
    E_1(t) &= e^{-\gamma_1t} E_1(0) + \sigma_{11}\int_0^t e^{-\gamma_1(t-s)} {\rm d}B_1(s) + \sigma_{12}\int_0^t e^{-\gamma_1(t-s)} {\rm d}B_2(s), \nonumber \\
    E_2(t) &= e^{-\gamma_2t} E_2(0) + \sigma_{21}\int_0^t e^{-\gamma_2(t-s)} {\rm d}B_1(s) + \sigma_{22}\int_0^t e^{-\gamma_2(t-s)} {\rm d}B_2(s). \nonumber
\end{align}
From this we compute the mean values
\begin{align}
    \left<E_1(t)\right>=\left<E_1(0)\right> e^{-\gamma_1 t}, \nonumber \\
    \left<E_2(t)\right>=\left<E_2(0)\right> e^{-\gamma_2 t}, \nonumber
\end{align}
as well as the correlation functions
\begin{align}
    {\rm Cov}\left[E_1(t),E_1(s)\right] &= \left< E_1(0)^2\right> e^{-\gamma_1(t+s)} + \frac{\sigma_{11}^2+\sigma_{12}^2+2\xi\sigma_{11}\sigma_{12}}{2\gamma_1} \left[e^{-\gamma_1|t-s|}-e^{-\gamma_1(t+s)}\right], \nonumber \\
    {\rm Cov}\left[E_2(t),E_2(s)\right] &= \left< E_2(0)^2\right> e^{-\gamma_2(t+s)} + \frac{\sigma_{21}^2+\sigma_{22}^2+2\xi\sigma_{21}\sigma_{22}}{2\gamma_2} \left[e^{-\gamma_2|t-s|}-e^{-\gamma_2(t+s)}\right], \nonumber \\
    {\rm Cov}\left[E_1(t),E_2(s)\right] &= \left<E_1(0),E_2(0)\right> e^{-\gamma_1t-\gamma_2s}+ \frac{\sigma_{11}\sigma_{21}+\sigma_{12}\sigma_{22}+\xi\sigma_{11}\sigma_{22}+\xi\sigma_{12}\sigma_{21}}{\gamma_1+\gamma_2} e^{-\gamma_1t-\gamma_2s} \left[e^{(\gamma_1+\gamma_2)\min(s,t)}-1\right], \nonumber \\
    {\rm Cov}\left[E_2(t),E_1(s)\right] &= \left<E_1(0),E_2(0)\right> e^{-\gamma_2t-\gamma_1s}+ \frac{\sigma_{11}\sigma_{21}+\sigma_{12}\sigma_{22}+\xi\sigma_{11}\sigma_{22}+\xi\sigma_{12}\sigma_{21}}{\gamma_1+\gamma_2} e^{-\gamma_2t-\gamma_1s} \left[e^{(\gamma_1+\gamma_2)\min(s,t)}-1\right]. \nonumber
\end{align}
Using these we find the spectral density matrix
for the correlated processes as 
\begin{equation}
    J(\omega) = \frac{1}{2\pi} 
    \begin{bmatrix} \frac{\sigma_{11}^2+2\xi\sigma_{11}\sigma_{22}+\sigma_{12}^2}{\gamma_1^2+\omega^2} & \frac{\sigma_{12}\sigma_{22}+\sigma_{11}\sigma_{21}+\xi(\sigma_{12}\sigma_{21}+\sigma_{11}\sigma_{22})}{(\gamma_1+i\omega)(\gamma_2-i\omega)} \\
      \frac{\sigma_{12}\sigma_{22}+\sigma_{11}\sigma_{21}+\xi(\sigma_{12}\sigma_{21}+\sigma_{11}\sigma_{22})}{(\gamma_1-i\omega)(\gamma_2+i\omega)} & \frac{\sigma_{21}^2+2\xi\sigma_{21}\sigma_{22}+\sigma_{22}^2}{\gamma_2^2+\omega^2}
    \end{bmatrix}.
\end{equation}

% !TEX root=Noise-enhanced-fidelity.tex

\section{Redfield tensor elements for cross correlation between $x$ and $z$  for a single $SU(2)$ qubit}.

The Bloch-Redfield equations
give the quantum dynamics of the
reduced density matrix according to 
\begin{align}
    d_t\rho_{\alpha\alpha'} = -i (\omega_\alpha-\omega_\alpha')\rho_{\alpha\alpha'}
    -\sum_{\beta\beta'}{\cal R}_{\alpha\alpha';\beta\beta'}(\rho_{\beta\beta'}-\rho_{beta\beta'}^{eq}
\end{align}
where  $\rho^{eq}$ is the equilibrium reduced density matrix and ${\cal R} $ is the Bloch-Redfield tensor
with elements 
\cite{Konrat:1993aa,redfield1957theory,Solomon1955559,ArgyresKelley1964}
\begin{align}
\begin{split}
     {\cal R}_{\alpha\alpha';\beta\beta'} 
    =
    \sum_{nm}\left\{
    \delta_{\alpha'\beta'}
    \sum_{\gamma}J_{nm}(\omega_\beta-\omega_\gamma)(A_n)_{\gamma\beta} (A_m)_{\alpha\gamma} \right.\\  
    \left.- (J_{nm}(\omega_\alpha'-\omega_\beta') + J_{nm}(\omega_\beta-\omega_\alpha))(A_n)_{\beta'\alpha'} (A_m)_{\alpha\beta} \right. \\
    +\left.\delta_{\alpha\beta}
    \sum_\gamma J_{nm}(\omega_\gamma-\omega_\beta')(A_n)_{\beta'\gamma} (A_m)_{\gamma \alpha'}
    \right\}
\end{split}
\label{eq:5}
\end{align}
where 
$(A_n)_{\alpha\beta} = \langle\alpha | \hat A_n | \beta\rangle$ are the matrix elements of the $\hat A_n$ operator in the 
eigenbasis of $H_o$ and 
$J_{nm}(\omega)$ are elements of the generalized spectral matrix 
characterizing the coupling between the 
system and its environment. 
Table ~\ref{tab:1} 
gives the tensor elements
for the case of a single qubit
with transition frequency $\epsilon$ 
coupled to a noisy environment through both longitudinal (through $\hat\sigma_z$) and transverse (through$\hat\sigma_x$ or $\hat\sigma_y$).
$J_{ij}(\omega)$ to denote the
spectral density associated
with the correlation function $\langle E_i(t)E_j(t')\rangle$.

The second column indicates whether or not $R_{ijkl}$ is non-vanishing within the
secular approximation. 
\begin{align}
    R^{sec}_{ij;kl} = \left(\delta_{ij}\delta_{kl} + \delta_{ik}\delta_{jl}(1-\delta_{ij}\delta_{kl})\right)
    R_{ij;kl}
\end{align}
When operating under this limit, the system populations are decoupled from the coherences, following a regular Pauli Master equation with the population rate matrix $R_{iikk}$. 
This ensures population conservation and achieves the correct thermal equilibrium over extended periods. 
Under this approximation, the density matrix exhibits the appropriate physical behaviour with ${\rm Tr}[\rho]=1$. The population rate matrix, being real, facilitates exponential relaxation of the populations. Coherences are also fully separated from the population and experience attenuation by the dephasing rates $R_{ij;ij}$. Generally, $R_{ij;ij}$ is complex, with its imaginary component representing bath-induced energy shifts.
For example, under the secular approximation, we expect that
$R_{11;11}+R_{22;11} = R_{22;22} + R_{11;22} =0$ and
$R_{12;12} = R_{21;21}^*$ for the coherence terms. 
The presence of the cross-correlation terms does not lead to a violation of these
conditions. 

\begin{table*}[h]
    \centering
    \begin{align}
    \begin{array}{ccccl|l|l}
    i& j&k&l &\text{Secular} &{\cal R}_{ij;kl}  & \text{Ornstein-Uhlenbeck}\\
    \hline
    \rowcolor{LightCyan}
 1 & 1 & 1 & 1 & \text{Secular} & J_{\text{xx}}(-\epsilon )+J_{\text{xx}}(\epsilon ) & \frac{2 s_x \left(s_x+2 \xi  s_{\text{xz}}\right)}{\gamma _x^2+\epsilon ^2} \\
 1 & 1 & 1 & 2 & \text{Non-Secular} & -J_{\text{xz}}(0)-J_{\text{zx}}(0) & -\frac{2 \left(s_x \left(s_{\text{xz}}+\xi  s_z\right)+\xi  s_{\text{xz}}^2\right)}{\gamma _x
   \gamma _z} \\
 1 & 1 & 2 & 1 & \text{Non-Secular} & J_{\text{xz}}(-\epsilon )-J_{\text{zx}}(-\epsilon )-2 J_{\text{zx}}(0) & -\frac{2 \left(\gamma _x^2 \left(i \epsilon  \gamma
   _z+\gamma _z^2+\epsilon ^2\right)-i \epsilon  \gamma _x \gamma _z^2+\epsilon ^2 \left(\gamma _z^2+\epsilon ^2\right)\right) \left(s_x \left(s_{\text{xz}}+\xi 
   s_z\right)+\xi  s_{\text{xz}}^2\right)}{\gamma _x \gamma _z \left(\gamma _x^2+\epsilon ^2\right) \left(\gamma _z^2+\epsilon ^2\right)} \\
   \rowcolor{LightCyan}
 1 & 1 & 2 & 2 & \text{Secular} & -J_{\text{xx}}(-\epsilon )-J_{\text{xx}}(\epsilon ) & -\frac{2 s_x \left(s_x+2 \xi  s_{\text{xz}}\right)}{\gamma _x^2+\epsilon ^2} \\
 1 & 2 & 1 & 1 & \text{Non-Secular} & -2 J_{\text{xz}}(-\epsilon )-J_{\text{xz}}(0)+J_{\text{zx}}(0) & -\frac{2 \left(s_x \left(s_{\text{xz}}+\xi  s_z\right)+\xi 
   s_{\text{xz}}^2\right)}{\left(\epsilon +i \gamma _x\right) \left(\epsilon -i \gamma _z\right)} \\
   \rowcolor{LightCyan}
 1 & 2 & 1 & 2 & \text{Secular} & 2 \left(J_{\text{xx}}(\epsilon )+2 J_{\text{zz}}(0)\right) & \frac{4 s_{\text{xz}}^2 \left(\gamma _x^2+\epsilon ^2\right)+4 \xi 
   s_{\text{xz}} \left(2 s_z \left(\gamma _x^2+\epsilon ^2\right)+s_x \gamma _z^2\right)+2 s_x^2 \gamma _z^2}{\gamma _z^2 \left(\gamma _x^2+\epsilon ^2\right)} \\
 1 & 2 & 2 & 1 & \text{Non-Secular} & -2 J_{\text{xx}}(-\epsilon ) & -\frac{2 s_x \left(s_x+2 \xi  s_{\text{xz}}\right)}{\gamma _x^2+\epsilon ^2} \\
 1 & 2 & 2 & 2 & \text{Non-Secular} & J_{\text{xz}}(-\epsilon )+J_{\text{zx}}(-\epsilon ) & \frac{2 \left(\gamma _x \gamma _z+\epsilon ^2\right) \left(s_x
   \left(s_{\text{xz}}+\xi  s_z\right)+\xi  s_{\text{xz}}^2\right)}{\left(\gamma _x^2+\epsilon ^2\right) \left(\gamma _z^2+\epsilon ^2\right)} \\
 2 & 1 & 1 & 1 & \text{Non-Secular} & -J_{\text{xz}}(\epsilon )-J_{\text{zx}}(\epsilon ) & -\frac{2 \left(\gamma _x \gamma _z+\epsilon ^2\right) \left(s_x
   \left(s_{\text{xz}}+\xi  s_z\right)+\xi  s_{\text{xz}}^2\right)}{\left(\gamma _x^2+\epsilon ^2\right) \left(\gamma _z^2+\epsilon ^2\right)} \\
 2 & 1 & 1 & 2 & \text{Non-Secular} & -2 J_{\text{xx}}(\epsilon ) & -\frac{2 s_x \left(s_x+2 \xi  s_{\text{xz}}\right)}{\gamma _x^2+\epsilon ^2} \\
 \rowcolor{LightCyan}
 2 & 1 & 2 & 1 & \text{Secular} & 2 \left(J_{\text{xx}}(-\epsilon )+2 J_{\text{zz}}(0)\right) & \frac{4 s_{\text{xz}}^2 \left(\gamma _x^2+\epsilon ^2\right)+4 \xi 
   s_{\text{xz}} \left(2 s_z \left(\gamma _x^2+\epsilon ^2\right)+s_x \gamma _z^2\right)+2 s_x^2 \gamma _z^2}{\gamma _z^2 \left(\gamma _x^2+\epsilon ^2\right)} \\
 2 & 1 & 2 & 2 & \text{Non-Secular} & 2 J_{\text{xz}}(\epsilon )+J_{\text{xz}}(0)-J_{\text{zx}}(0) & \frac{2 \left(s_x \left(s_{\text{xz}}+\xi  s_z\right)+\xi 
   s_{\text{xz}}^2\right)}{\left(\epsilon -i \gamma _x\right) \left(\epsilon +i \gamma _z\right)} \\
   \rowcolor{LightCyan}
 2 & 2 & 1 & 1 & \text{Secular} & -J_{\text{xx}}(-\epsilon )-J_{\text{xx}}(\epsilon ) & -\frac{2 s_x \left(s_x+2 \xi  s_{\text{xz}}\right)}{\gamma _x^2+\epsilon ^2} \\
 2 & 2 & 1 & 2 & \text{Non-Secular} & -J_{\text{xz}}(\epsilon )+J_{\text{zx}}(\epsilon )+2 J_{\text{zx}}(0) & \frac{2 \left(\gamma _x^2 \left(-i \epsilon  \gamma
   _z+\gamma _z^2+\epsilon ^2\right)+i \epsilon  \gamma _x \gamma _z^2+\epsilon ^2 \left(\gamma _z^2+\epsilon ^2\right)\right) \left(s_x \left(s_{\text{xz}}+\xi 
   s_z\right)+\xi  s_{\text{xz}}^2\right)}{\gamma _x \gamma _z \left(\gamma _x^2+\epsilon ^2\right) \left(\gamma _z^2+\epsilon ^2\right)} \\
 2 & 2 & 2 & 1 & \text{Non-Secular} & J_{\text{xz}}(0)+J_{\text{zx}}(0) & \frac{2 \left(s_x \left(s_{\text{xz}}+\xi  s_z\right)+\xi  s_{\text{xz}}^2\right)}{\gamma _x
   \gamma _z} \\
   \rowcolor{LightCyan}
 2 & 2 & 2 & 2 & \text{Secular} & J_{\text{xx}}(-\epsilon )+J_{\text{xx}}(\epsilon ) & \frac{2 s_x \left(s_x+2 \xi  s_{\text{xz}}\right)}{\gamma _x^2+\epsilon ^2} \\
 \hline
 \end{array}
    \nonumber
\end{align}
    \caption{Redfield tensor elements for $SU(2)$ qubit with 
    cross-correlation between $\hat\sigma(1)$ and $\hat\sigma(3)$ noise terms. The second column indicates whether the term survives under the secular approximation, which separates the evolution
of the population and the coherence terms.
    The last column gives the tensor element within the Ornstein-Uhlenbeck model.
    % generated in the "Redfield+spectral_density.nb"
    }
    \label{tab:1}
\end{table*}

\end{widetext}

\end{document}